\documentclass[
prd,
preprint,
superscriptaddress,s
showpacs,
showkeys,
nofootinbib,
aps,
amsfonts,
amssymb,
]{revtex4-1}

\usepackage{color}
\usepackage{multirow} 
\usepackage{amsmath}
\usepackage{graphicx}
\usepackage{slashed}
\usepackage{epic}
\usepackage{eepic}
\usepackage{epsfig}
\usepackage{latexsym}
\usepackage{float}
\usepackage{amsfonts}
\usepackage{amssymb} 
\usepackage{color}
\usepackage{multirow,array}
\usepackage{tikz}
\usepackage[colorlinks, urlcolor=blue, citecolor=red, link
color=blue]{hyperref}

\usepackage[caption=false]{subfig}

\usepackage{geometry}
\newcommand{\lsim}{\mbox{\raisebox{-.6ex}{~$\stackrel{<}{\sim}$~}}}
\newcommand{\gsim}{\mbox{\raisebox{-.6ex}{~$\stackrel{>}{\sim}$~}}}

\newcolumntype{L}[1]{>{\raggedright\arraybackslash}p{#1}}
\newcolumntype{C}[1]{>{\centering\arraybackslash}p{#1}}
\newcolumntype{R}[1]{>{\raggedleft\arraybackslash}p{#1}}
                                                                          
\begin{document}

\preprint{IPPP/16/91}

\title{Type-II Seesaw and Multilepton Signatures at Hadron Colliders}

\newcommand{\iiserm}{\affiliation{Department of Physics, Indian Institute of Science Education and Research Mohali (IISER Mohali), Sector 81, SAS Nagar, Manauli 140306, India}}
\newcommand{\du}{\affiliation{Department of Physics and Astrophysics, University of Delhi, New Delhi 110007, India}}
\newcommand{\ippp}{\affiliation{Institute for Particle Physics Phenomenology {(IPPP)}, Department of Physics, Durham University, Durham, DH1 3LE, UK}}

\author{Manimala Mitra} \iiserm \author{Saurabh Niyogi} \du \author{Michael Spannowsky} \ippp

\begin{abstract}{We investigate multilepton signatures, arising from the decays of doubly charged and singly charged Higgs bosons in the Type-II Seesaw model. Depending on the vacuum expectation value of the triplet $v_{\Delta}$,
the doubly and singly charged Higgs bosons can decay into a large variety of multi-lepton final states. We explore all possible decay modes corresponding to different regimes of $v_{\Delta}$, that generate distinguishing four and five leptonic signatures. We focus on the 
13 TeV Large Hadron Collider (LHC) and further extend the study to a very high energy proton-proton collider (VLHC) with a center-of-mass energy of 100 TeV. We find that a doubly charged Higgs boson of masses around 375 GeV can be discovered at immediate LHC runs. A heavier mass of 630 GeV can instead be discovered at the high-luminosity run of the LHC or at the VLHC with 30 $\rm{fb}^{-1}$. }


\end{abstract}


\maketitle

\section{Introduction}
The observation of nonzero neutrino masses and their  mixings provide unambiguous experimental evidence of physics beyond the SM (BSM). So far, oscillation experiments have measured the solar and atmospheric  
mass square differences $\Delta m^2_{12}$, $|\Delta m^2_{13}|$ and the mixing angles $\theta_{12}$, $\theta_{23}$ and $\theta_{13}$ \cite{Gonzalez-Garcia:2015qrr}. Additionally, the cosmological constraints on the sum of light neutrino masses \cite{Ade:2015xua} bound the SM neutrinos to be less than eV. A natural explanation for small  neutrino masses is provided by the seesaw mechanism, where eV-scale neutrino masses are generated from lepton number violating (LNV) operators of dimension 5~\cite{Weinberg:1979sa,Wilczek:1979hc}. UV complete models generating the higher dimensional  operator can have a right handed neutrino $N_R$ (Type-I seesaw) \cite{Minkowski:1977sc,Mohapatra:1979ia,Yanagida:1979as,GellMann:1980vs,Schechter:1980gr,Shrock:1980ct},
$SU(2)_L$ triplet Higgs $\Delta_L$ (Type-II seesaw) \cite{Magg:1980ut,Cheng:1980qt,Lazarides:1980nt,Mohapatra:1980yp}, or, a $SU(2)_L$ triplet fermionic field $\Sigma$ (Type-III seesaw) \cite{Foot:1988aq}. The Type-I and Type-II seesaw models can further be embedded into Left-Right symmetric models \cite{LR}. 
Another very popular model is the inverse seesaw model \cite{invo, inverseso10} where  the light neutrino masses are proportional to a small LNV parameter, and thus tend to zero for a vanishingly small value of the parameter. In this model, the  smallness of the neutrino mass is protected by the lepton number symmetry of the Lagrangian\footnote{For reviews of TeV-scale seesaw models and their phenomenology see Refs.~\cite{Chen:2011de}.}.  If we have 
a right handed neutrino or, a Higgs triplet states with low masses (few hundreds GeV upto a few TeV), these BSM states can be directly produced at the LHC and can be detected via their decay products in direct searches \cite{KS,Chen:2013fna, Mitra:2016kov, Perez:2008ha,Melfo:2011nx,delAguila:2008cj, Chakrabarti:1998qy, Aoki:2011pz, Chun:2013vma, Akeroyd:2005gt, Banerjee:2013hxa, Dev:2013ff}.
Apart from direct production, these states may appear in loop for various processes/decays. Strong constraints can be put on doubly and singly charged scalar masses
from Lepton Number Violating processes at the LHC \cite{delAguila:2013mia,delAguila:2013aga}\footnote{For further discussion on the recent developments on muon anomalous magnetic moment and Lepton flavor violation in the context of several extensions of the SM, see review \cite{Lindner:2016bgg}.}.
Another way to infer the existence of such resonances are indirect detection experiments which  also cover a wide range of masses and mixings \cite{MSV}. 

Here we focus on the Type-II seesaw mechanism. The model  is augmented with a doubly charged Higgs boson that can give rise to the smoking gun signal of same-sign dilepton pairs \cite{Muhlleitner:2003me,Perez:2008ha, Melfo:2011nx, delAguila:2008cj}. The neutral component of the triplet Higgs develops a vacuum expectation value (vev) $v_{\Delta}$, 
and generates neutrino masses through the Yukawa Lagrangian. In addition to the doubly charged Higgs, the model also contains a singly charged Higgs state. The 
details of the Higgs spectrum have been discussed in \cite{Arhrib:2011uy,Dev:2013ff}. The particles' branching ratios and some collider signatures have been outlined in \cite{Perez:2008ha,Melfo:2011nx,delAguila:2008cj}. CMS and ATLAS  have searched for pair production of doubly charged Higgs bosons, 
followed by their decay into same-sign dileptons and, hence, set a limit on the mass of $H^{++}$ \cite{atlashpp}.
An alternative search where the $H^{++}$ is produced in association with two jets was found to be less constraining \cite{Khachatryan:2014sta}.  
Other than the dileptonic decay mode, in parts of the parameter space with relatively large triplet vev $v_{\Delta}$, the charged Higgs dominantly decays into gauge bosons or via cascade decays with on/off-shell $W$ bosons \cite{Chakrabarti:1998qy,Melfo:2011nx,Aoki:2011pz}. The later can give 
rise to the distinctive same-sign dilepton signatures, with additional b-jets \cite{Aoki:2011pz}. In particular, same-sign diboson scenario has been studied in the 
context of LHC in \cite{Kanemura:2013vxa,Kanemura:2014goa,Kanemura:2014ipa}.

Other than the conventional channel of pair production of $H^{++}$ that leads to four leptonic final states, one can have even up to five or six leptons via cascade decays into gauge bosons\footnote{Multilepton signature with triplet Higgs in supersymmetric scenario has also been studied in \cite{Bandyopadhyay:2014vma} where singly charged Higgs
decays into two gauge bosons.}. The multileptonic final states provide very clean signatures at hadron colliders. Hence, a handful of events can confirm or rule out the model.
In this work, we carry out a thorough investigation on the collider search of such multilepton (four or five leptons) final states, that will be useful to probe the complete range of $v_{\Delta}=10^{-9}$~GeV$-1$~GeV. We divide this range into three regimes: small, intermediate and large $v_{\Delta}$.
We focus on both, the immediate run-II of the LHC with 13 TeV center-of-mass energy and also its future high-luminosity upgrade (HL-LHC). 
We further analyse the detection prospects of such multilepton signatures at a possible future $100$ TeV proton-proton collider (Very Large Hadron Collider (VLHC)) \cite{Arkani-Hamed:2015vfh,Golling:2016gvc}.


Our paper is organized as follows: we briefly review the basics of the Type-II seesaw model in Section~\ref{model}. In Sec.~\ref{dcbr}, we discuss the relevant decay modes and 
branching ratios. In the subsequent sections, Sec.~\ref{prod} and Sec.~\ref{analyses}, we analyse in detail the production cross sections and the discovery potential of 
the multilepton final states. Finally, we present our conclusion in Sec.~\ref{conclu}.

\section{Model \label{model}}
In this section, we briefly review the basics of the Type-II seesaw scenario \cite{Magg:1980ut,Cheng:1980qt,Lazarides:1980nt,Mohapatra:1980yp}. The model consists of the SM fields, with 
a Higgs doublet $\Phi$ and an additional  $SU(2)_L$ triplet  Higgs $\Delta$ that has hypercharge $U(1)_Y=2$: 
\begin{eqnarray}
\Phi= \begin{pmatrix} \Phi^+ \\ \Phi^0  \end{pmatrix} 
  \, \, \, \, \,~~~ ~~~\rm{and} ~~~~~~
 \Delta=\begin{pmatrix} \frac{\Delta^+}{\sqrt{2}} & \Delta^{++} \\ \Delta^0 & -\frac{\Delta^+}{\sqrt{2}}
\end{pmatrix}.
\end{eqnarray}
The neutral components of the doublet and triplet Higgs fields are  $\Phi^0=\frac{1}{\sqrt{2}}(\phi^0+i\chi^0)$ and $\Delta^0=\frac{1}{\sqrt{2}}(\delta^0+i\eta^0)$ respectively.
The components $\phi^0$ and $\delta^0$ develop a vev denoted as $v_{\Phi}$ and $v_{\Delta}$ with the light neutrino masses $m_{\nu} $ being proportional to the triplet vev $v_{\Delta}$. The two vevs satisfy $v^2=v^2_{\Phi}+v^2_{\Delta}=(246 \, \, \rm{GeV})^2$. The kinetic term of the new scalar field $\Delta$ that generates the interactions with the SM gauge bosons, has the form
\begin{eqnarray}
\mathcal{L}_{\rm{kin}}( \Delta)&=&\rm{Tr}[(D_\mu \Delta)^\dagger (D^\mu \Delta)].
\label{kinetic}
\end{eqnarray}
The covariant derivative of Eq.~(\ref{kinetic}) is defined as 
\begin{eqnarray}
D_\mu \Delta=\partial_\mu \Delta+i\frac{g}{2}[\tau^aW_\mu^a,\Delta]+ig'B_\mu\Delta.
\label{covtrip} 
\end{eqnarray}
In addition, $\Delta$ also interacts with the leptons through the Yukawa interaction
\begin{eqnarray}
\mathcal{L}_Y(\Phi, \Delta)&=& Y_{\Delta}\overline{L_L^{c}}i\tau_2\Delta L_L+\rm{h.c.}.~~~~ 
\label{yukawa}
\end{eqnarray}
Here, $c$ represents the charge conjugation transformation, and $Y_{\Delta}$ is a $3\times 3$ matrix. The triplet field $\Delta$ 
carries lepton number +2 and hence the Yukawa term conserves lepton number. The scalar potential of the Higgs fields $\Phi$ and $\Delta$ is  
\begin{eqnarray}
V(\Phi,\Delta)&=&m_\Phi^2\Phi^\dagger\Phi+\tilde{M}^2_{\Delta}\rm{Tr}(\Delta^\dagger\Delta)+\left(\mu \Phi^Ti\tau_2\Delta^\dagger \Phi+\rm{h.c.}\right)+\frac{\lambda}{4}(\Phi^\dagger\Phi)^2 \nonumber\\
&+&\lambda_1(\Phi^\dagger\Phi)\rm{Tr}(\Delta^\dagger\Delta)+\lambda_2\left[\rm{Tr}(\Delta^\dagger\Delta)\right]^2 +\lambda_3\rm{Tr}[(\Delta^\dagger\Delta)^2]
+\lambda_4\Phi^\dagger\Delta\Delta^\dagger\Phi,~~~~
\label{eqn:scalpt}
\end{eqnarray}
where $m_{\Phi}$ and $\tilde{M}_{\Delta}$ are real parameters with mass dimension 2, $\mu$ is the lepton-number violating
parameter with positive mass dimension and $\lambda$, $\lambda_{1-4}$ are dimensionless quartic Higgs couplings.
Minimization conditions for each of the scalar fields can be used to replace any two of the parameters\footnote{For the discussion on the  minimization of the scalar potential, see \cite{Arhrib:2011uy}.}. Usually the two mass
parameters $m_{\Phi}^2$ and $\tilde{M}^2_{\Delta}$ are eliminated which leaves six independent parameters.

After transforming into the mass eigenbasis the two charged scalar fields
$\Phi^{\pm}$ and $\Delta^{\pm}$ mix to the charged Higgs bosons $\chi^{\pm}$ and $H^{\pm}$ \cite{Aoki:2011pz}. Similarly, the mixing between the two CP-odd fields ($\chi^{0}$ and $\eta^{0}$)
gives rise to $\rho^{0}$ and $A$. Finally, we obtain the SM Higgs field ($h$) and a heavy Higgs ($H$) by mixing the two neutral CP-even states $\Phi^{0}$ and $\delta^{0}$. $\chi^\pm$ and $\rho^{0}$ act as the three Goldstone bosons which give masses to the SM weak gauge bosons. The remaining seven states are the 
physical Higgs bosons.

%


The masses of {the} doubly and singly charged Higgs states $H^{\pm \pm}$ and $H^{\pm}$  are expressed in terms of the parameters in the Lagrangian as 
\begin{eqnarray}
m_{H^{++}}^2&=&M_\Delta^2-v_\Delta^2\lambda_3-\frac{\lambda_4}{2}v_\Phi^2,\label{eq:mhpp}\\
m_{H^+}^2&= &\left(M_\Delta^2-\frac{\lambda_4}{4}v_\Phi^2\right)\left(1+\frac{2v_\Delta^2}{v_\Phi^2}\right).\label{eq:mhp}
\end{eqnarray}
The CP-even and CP-odd neutral Higgs $H$, $h$ and $A$ have the following masses:
\begin{eqnarray}
m_h^2&=&\mathcal{T}_{11}^2\cos^2\alpha+\mathcal{T}_{22}^2\sin^2\alpha-\mathcal{T}_{12}^2\sin2\alpha, \label{mh}\\ 
m_H^2&=&\mathcal{T}_{11}^2\sin^2\alpha+\mathcal{T}_{22}^2\cos^2\alpha+\mathcal{T}_{12}^2\sin2\alpha,\label{mH}\\
m_A^2 &= &M_\Delta^2\left(1+\frac{4v_\Delta^2}{v_\Phi^2}\right) \label{mA},
\end{eqnarray}
where $\mathcal{T}_{11}^2$, $\mathcal{T}_{22}^2$ and $\mathcal{T}_{12}^2$ are given by \cite{Aoki:2011pz}, 
\begin{eqnarray}
\mathcal{T}_{11}^2&=&\frac{v_\Phi^2\lambda}{2},\\
\mathcal{T}_{22}^2&=&M_\Delta^2+2v_\Delta^2(\lambda_2+\lambda_3), ~~\quad M^2_{\Delta}=\frac{v^2_{\Phi} \mu}{\sqrt{2} v_{\Delta}}\\
\mathcal{T}_{12}^2&=&-\frac{2v_\Delta}{v_\Phi}M_\Delta^2+v_\Phi v_\Delta(\lambda_1+\lambda_4).
\end{eqnarray}

Note that, the difference between $H^{\pm \pm}$ and $H^{\pm}$ is proportional to the coupling $\lambda_4$, i.e.~ 
\begin{equation}
M^2_{H^{\pm}}-M^2_{H^{\pm \pm}} \sim \frac{\lambda_4}{2} v^2_{\Phi}+\mathcal{O}(v^2_{\Delta}).
\label{diffchdmass}
\end{equation}
\begin{figure}[t]
\centering
\subfloat[]{\includegraphics[width=8.00cm,height=6.0cm]{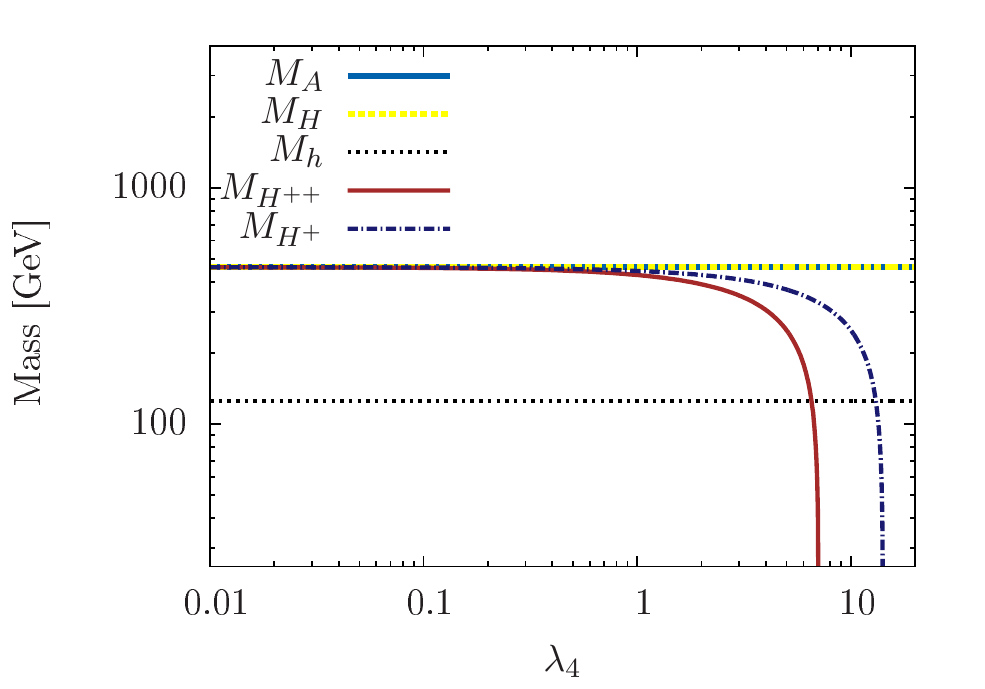}\label{fig:massspec1a}}~~
\subfloat[]{\includegraphics[width=8.00cm,height=6.0cm]{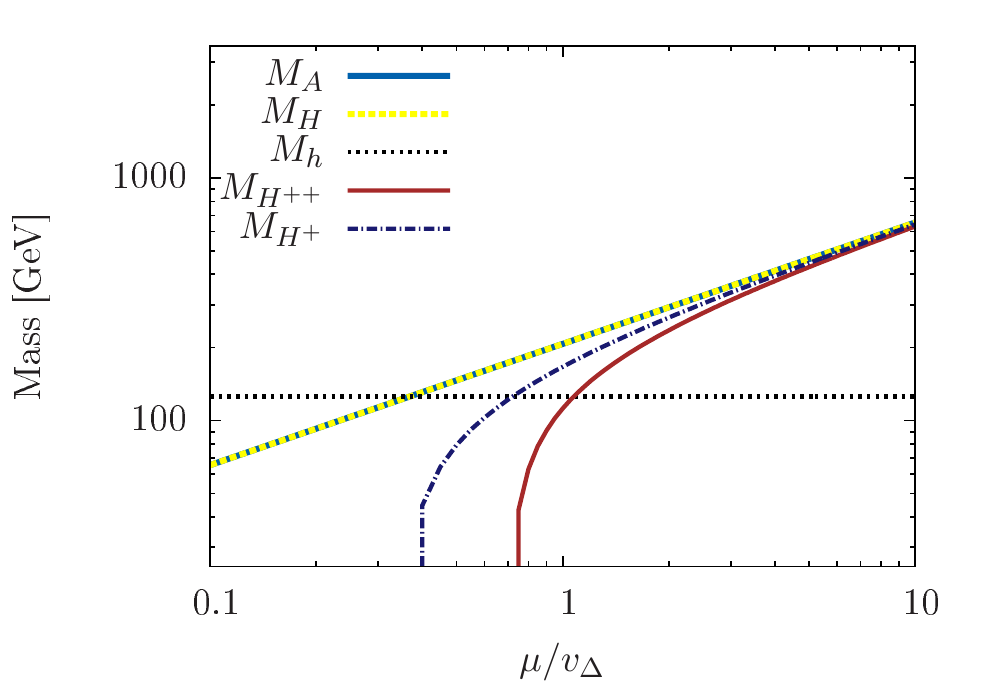}\label{fig:massspec1b}}
\caption{Variation of the masses of the Higgs states with the coupling $\lambda_4$ (left panel) and with $\mu/v_{\Delta}$ (right panel). The other parameters have been set to
$\lambda_{1,2,3}=1.0$, $\lambda=0.52$, and $v_{\Delta}=10^{-5}$ GeV. For the figure in the left panel, $\mu=5 \times 10^{-5}$ GeV and for right  panel $\lambda_4=1.0$. The mass of the SM Higgs is 125 GeV.}
\label{fig:massspec}
\end{figure}

The Higgs triplet $\Delta$  contributes to the gauge boson masses through its vacuum expectation value $v_{\Delta}$. The measurement of the $\rho$-parameter severely constrains the vev to  $v_{\Delta} \lsim 5$ GeV \cite{Kanemura:2012rs}. 
Since, in our case, $v_{\Delta}\ll v_{\Phi}$, the difference between the charged Higgs masses $M_{H^{\pm \pm}}$ and $M_{H^{\pm}}$ is
governed by the electroweak vev $v_{\Phi}$. In Fig.~\ref{fig:massspec1a}, we show the  mass spectrum of all the Higgs states, assuming 
the other parameters $\lambda_1=\lambda_2=\lambda_3=1.0$,  $\lambda=0.52$ and $\lambda_4 >0$. Note that, {for our choice of parameters}, the CP-odd state $A$ and 
the CP-even states $H$ are heavier than the charged Higgs states $H^{\pm}, H^{\pm \pm}$. Between the charged Higgs states, 
$H^{\pm}$ is heavier than $H^{\pm \pm}$, {as $\lambda_4 > 0$}. Finally, $h$ is the SM-like Higgs boson, assumed to have a mass of 125 GeV. The other regime $\lambda_4 < 0$
gives the opposite hierarchy between the charged Higgs masses, and has been explored in  \cite{Aoki:2011pz}. We show the variation of 
mass spectrum of the different Higgs states  with the ratio $\frac{\mu}{v_{\Delta}}$ in Fig.~\ref{fig:massspec1b}. 

It is evident from Fig.~\ref{fig:massspec1a}  that for $\lambda_4 \lesssim \mathcal{O}(0.1)$ all the Higgs bosons are almost degenerate in masses. 
For larger $\lambda_4$, the charged Higgs bosons $H^{\pm}$, $H^{\pm \pm}$  become lighter than the neutral Higgs states $H^0, A^0$. 
Higher $\lambda_4$ (well within the perturbative regime) results in a splitting between singly charged and doubly charged Higgs masses. 
In the subsequent analyses, we focus on the charged Higgses with masses avoiding LEP/LHC bounds and analyse their decay widths, branching fractions 
and collider signatures. Note that, both the charged Higgses $H^{\pm \pm}$ and $H^{\pm}$ contribute to $h \rightarrow \gamma \gamma$ process at one loop \cite{Arhrib:2011vc,Arhrib:2014nya,Das:2016bir}. Very low 
masses can cause deviation in the measured signal strength $\mu_{\gamma \gamma}$ at the LHC \cite{atlas-cms-mu-gammagamma}. Since no direct bound exists from LHC on these masses for 
$v_{\Delta} \gtrsim 10^{-5}$ GeV, it might be, therefore, possible to set lower bound on the masses of charged scalars from the observed value of the diphoton signal strength.
As mentioned in \cite{Das:2016bir}, the lower bounds on the charged scalar masses can be considered as $m_{H^{\pm}} \sim 130$ GeV and $m_{H^{\pm \pm}} \sim 100$ GeV allowing $2\sigma$ deviation from the central values of 
the T-parameter and observed diphoton signal strength.


\section{Decay Widths and Branching Ratios \label{dcbr}}
\begin{figure}[t]
\centering
\subfloat[]{\includegraphics[width=8.00cm,height=5.0cm]{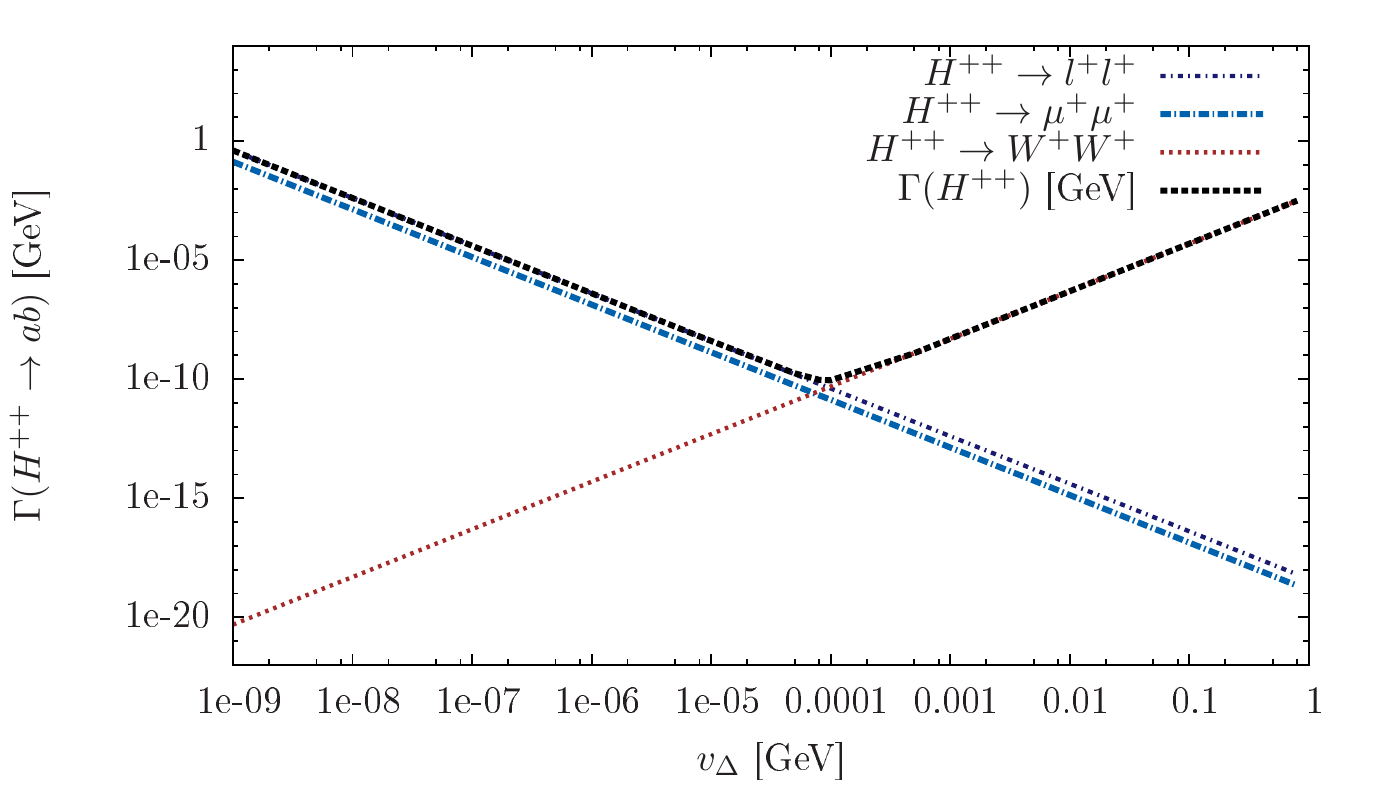}\label{fig:tw630}}~~
\subfloat[]{\includegraphics[width=8.0cm,height=5.0cm]{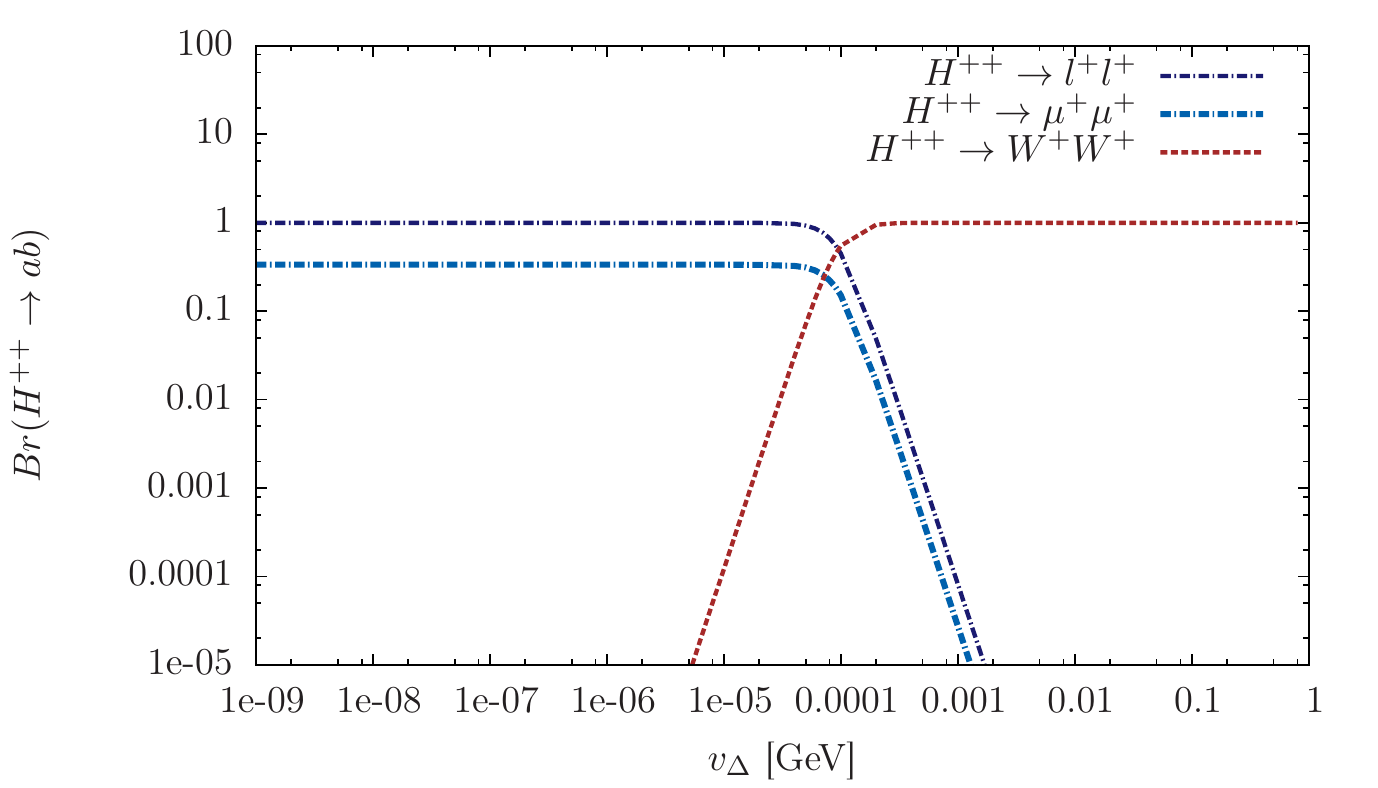}\label{fig:hpp630}}
\caption{Decay widths and branching ratios of the doubly charged Higgs $H^{++}$ into different final states. The dark (light) blue dot-dashed 
line represent the branching ratio of $H^{++} \to l^{+} l^{+} (\mu^{+} \mu^{+})$ states. The red dashed line represents
BR($H^{++} \to W^{+} W^{+} $). The masses of the Higgs triplet is $M_{H^{++}}= 630$ GeV. The total width is denoted by the black line.}
\label{fig:hppbr}
\end{figure}

\begin{figure}[h]
\centering
\subfloat[]{\includegraphics[width=9.0cm,height=6.0cm]{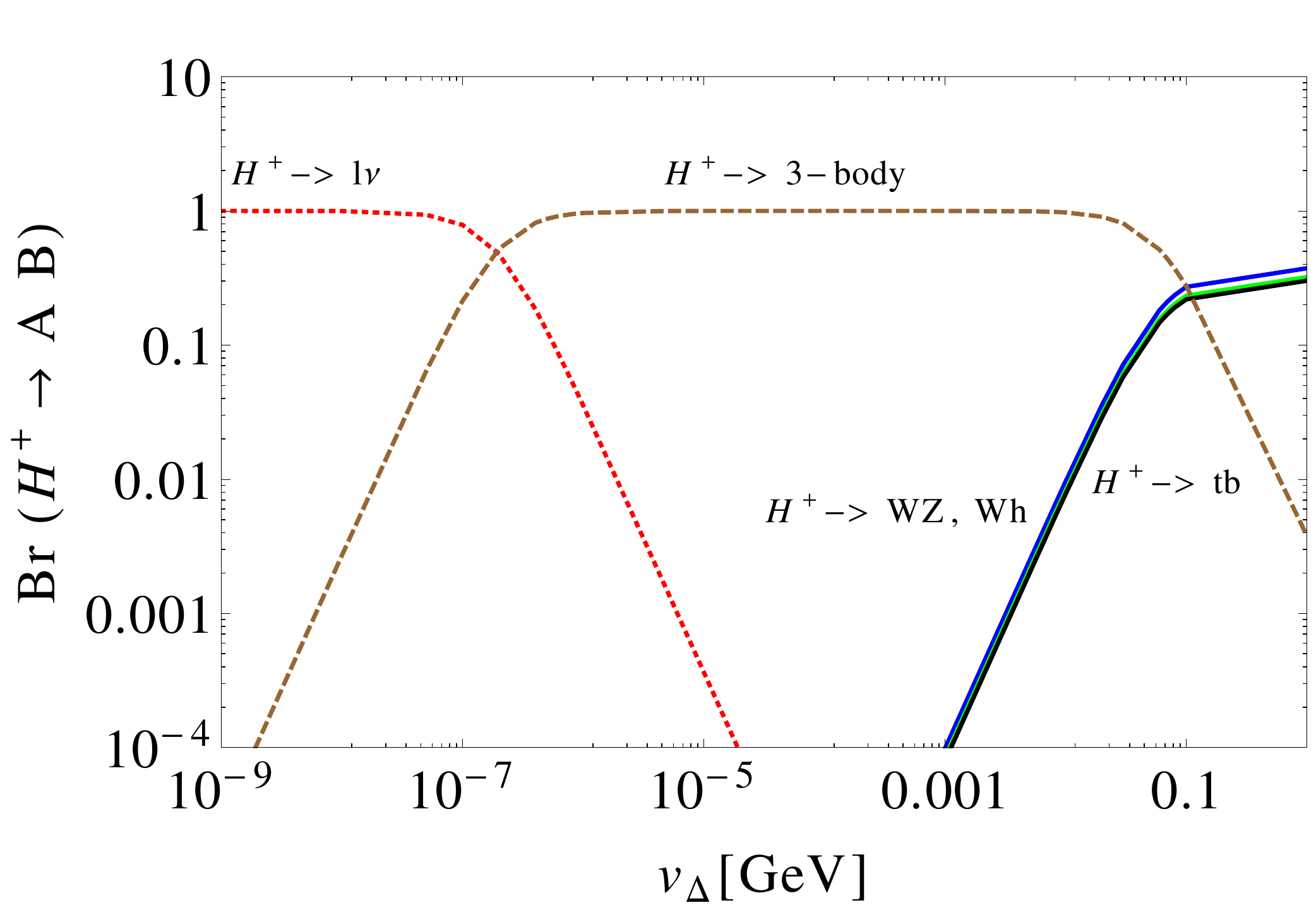}}
\caption{ {Branching ratios of the different decay modes of $H^{+}$, for the illustrative mass point,  $M_H^{+}=445$ GeV,  $M_H^{++}=428$ GeV. }}
\label{fig:hpbr}
\end{figure}

As outlined in the previous section, we consider a mass spectrum, where the singly charged Higgs is heavier than the doubly charged Higgs,  $M_{H^+} > M_{H^{++}}$. The doubly charged Higgs $H^{\pm \pm}$ has only a limited number of decay modes, i.e. decays into same-sign dileptons and same-sign dibosons. The partial decay widths of the dileptonic channel depends on the strength of the corresponding Yukawa coupling $Y_{\Delta}  \sim \frac{M_{\nu}}{v_{\Delta}}$. For small $v_{\Delta} \sim 10^{-9}$ GeV, this gives rise to large coupling strength $Y_{\Delta} \sim \mathcal{O}(1)$ and hence, a large partial width. The bosonic decay mode $H^{\pm \pm} \to W^{\pm }W^{\pm}$ is, on the other hand, proportional to the $v^2_{\Delta}$. Therefore, the partial width (and, also the branching ratio) for this channel becomes large for a large $v_{\Delta}$ \cite{Perez:2008ha, Melfo:2011nx}.
 
With the choice for our mass spectrum,
$H^{\pm\pm} \to H^{\pm} W^{\pm}$ remains absent. We show the decay widths and branching ratios in Fig.~\ref{fig:hppbr}, for the illustrative mass point $M_{H^{++}} \simeq 630$ GeV. This value is in agreement with current limits from ATLAS \cite{atlashpp}. The variation of the decay width and branching fractions with the doubly charged Higgs mass is nominal. Hence, we do not show them explicitly. A couple of comments are in order:
\begin{itemize}
\item
Leptonic final states are the dominant decay modes for smaller $v_{\Delta} \lsim 10^{-4}$ GeV. For $v_{\Delta} \gsim 10^{-4}$ GeV, $H^{++} \to W^{+}W^{+}$ becomes dominant, as evident from Fig.~\ref{fig:hppbr}, exceeding $H^{++} \to l^{+}l^{+}$
\cite{Perez:2008ha}.  
\item
The decay widths and the branching ratio of  $H^{++} \to l^+ l^+$ depends on the neutrino oscillation parameters $\theta_{12}, \theta_{23}, \theta_{13}$, the  light neutrino masses
$m_i$, and the CP violating phases. In our analysis we consider the best fit value of the oscillation parameters \cite{Gonzalez-Garcia:2015qrr}, $\theta_{12}=33.48^{\circ}, \theta_{13}=8.50^{\circ}, \theta_{23}=42.4^{\circ}$ and the light neutrino masses $m_1=0.10$ eV, $m_2=0.100376$ eV and $m_3=0.110589$ eV. We choose the CP-phases to be zero.  In Fig.~\ref{fig:hppbr}, the leptonic mode involves all three leptons $e, \mu,\tau$ and we separately show the $\mu \mu$ channel.

\item
The branching ratio of $H^{\pm \pm} \to \mu^{\pm} \mu^{\pm}$ is 31.5$\%$ for small $v_{\Delta} \lsim 10^{-5}$ GeV. The branching ratio to $ee$ and $\tau \tau$ are also comparable, while the $e \mu$, $e \tau$ and other 
off-diagonal branching ratios are relatively smaller.
 
\end{itemize} 

The singly charged Higgs, on the other hand, can decay to a number of final states, including $l^+ \nu$, $W^+Z$, $W^+h$, $W^+ H^{++}$ and $t\bar{b}$. 
For $M_{H^{+}} - M_{H^{++}} < M_{W^{+}}$, $H^{+}$ will also decay via the off-shell mode $H^{++} {W^{-}}^*$. We show the branching ratios of the different decay modes in Fig.~\ref{fig:hpbr} for the scenario $M_{H^{+}}-M_{H^{++}} < M_W$. It is evident from the left panel that for
$v_{\Delta} < 10^{-7}$ GeV, the leptonic mode $H^{+} \to l^{+} \nu$ ($l=e,\mu,\tau$) is the dominant decay channel. In the intermediate range of
$v_{\Delta} \sim 10^{-6}-10^{-2}$ GeV, $H^{+} \to H^{++} {W^{-}}^* \to H^{++} jj+l \nu$  is maximised. A similar feature of the branching ratios is present for the on-shell mode $H^{+} \to H^{++} W^{-}$.
Interestingly, for this case, the above mode remains dominant even for much lower values of $v_{\Delta} \sim 10^{-8}$ GeV.  Note that, although we have only shown the branching ratios for few illustrative mass points of $H^{+}$ and $H^{++}$, the features remain unaltered for other masses as well. This intermediate $v_{\Delta}$ region is of particular interest,
as the decay mode $H^{+} \to H^{++} W^{-}/H^{++} {W^{-}}^*$ can give distinctive multilepton signatures (with five/six leptons in the final state) at the LHC (and VLHC). We will explore this in detail in Sec.~\ref{intvdelta}.

\section{Production cross section at LHC and VLHC \label{prod}}
\begin{figure}[t]
\subfloat[]{\includegraphics[width=8.0cm,height=5.0cm]{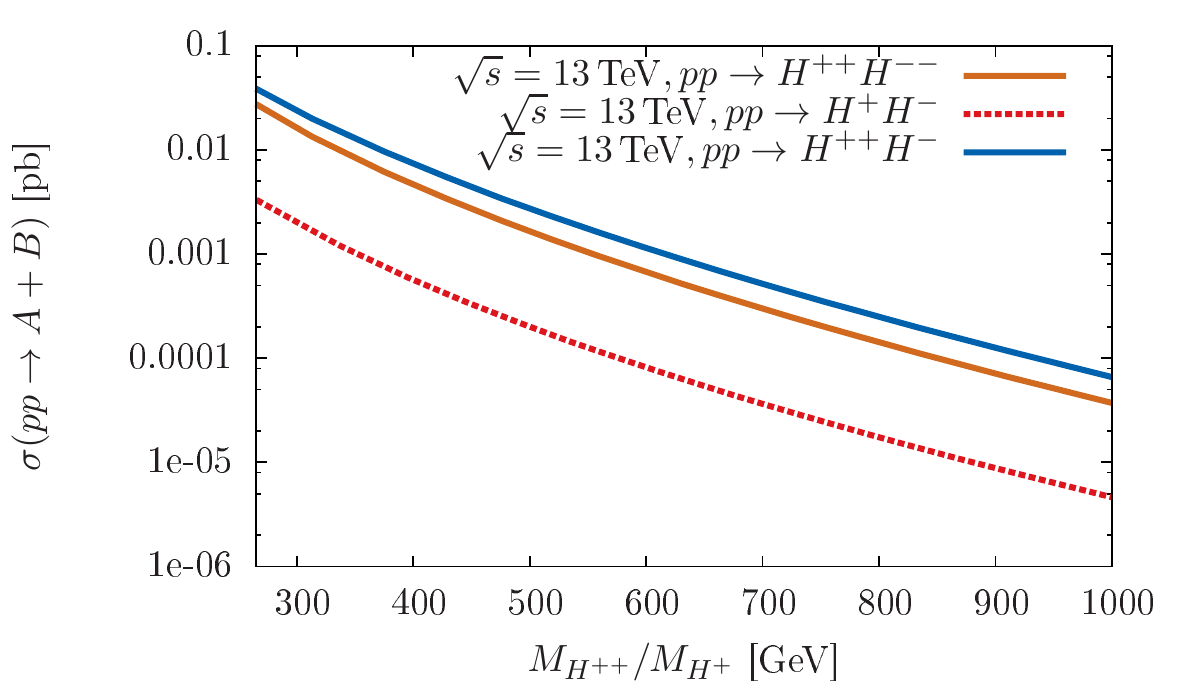}\label{fig:prod13}}~~
\subfloat[]{\includegraphics[width=8.0cm,height=5.0cm]{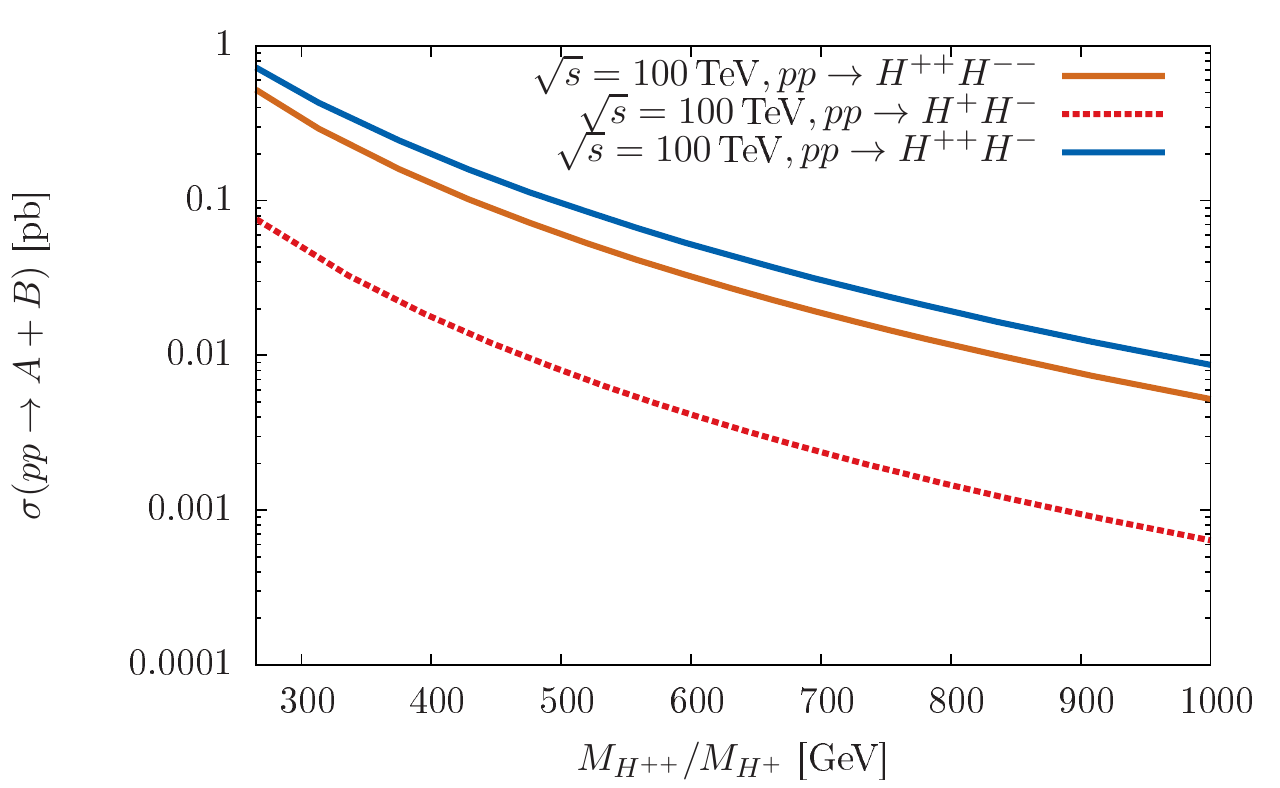}\label{fig:prod100}}
\caption{Variation of the production cross section of charged Higgs with mass. Left panel: 13 TeV, right panel: 100 TeV. The K-factor for 13 TeV  has been chosen as 1.25  \cite{Akeroyd:2005gt}. }
\label{fig:prodchd}
\end{figure}

In this section we discuss the production of the charged Higgs boson states at the LHC at 13 TeV and the VLHC with center-of-mass energy of 100 TeV.
The dominant processes are pair production of $H^{\pm \pm}/H^{\pm}$ through s-channel $Z/\gamma$ exchange and 
associated production of singly and doubly charged Higgs bosons, i.e. $pp \to H^{\pm \pm} H^{\mp}$, mediated by the $W$ boson.

We use  {\tt FeynRules} \cite{Alloul:2013bka} to generate suitable model file via Universal Feynrules Output (UFO) \cite{Degrande:2011ua, deAquino:2011ub} interface and compute the hard process with {\tt MadGraph5\_aMC@NLO} \cite{Alwall:2014hca}. Eventually, the generated events are showered and hadronised using {\tt Pythia} \cite{Sjostrand:2001yu,Sjostrand:2007gs}. To mimic the detector response, a fast detector simulation is performed  using {\tt Delphes-3.3} \cite{deFavereau:2013fsa}.
All cross sections have been evaulated with {\tt NN23LO1} \cite{Ball:2013hta} as parton distribution function.
We show the production cross section for the processes $p p \to H^{++} H^{--}$, $p p \to H^{+} H^{--}+h.c.$ and $H^{+} H^{-}$  in Fig.~\ref{fig:prod13}
and Fig.~\ref{fig:prod100} for the center-of-mass (c.o.m.) energies  $\sqrt{s}=13$ and $100$ TeV, respectively. For the 13 TeV c.o.m. energy we multiply the LO cross section by the K-factor K=1.25 \cite{Akeroyd:2005gt}.  As mentioned earlier, the production of $H^{++} H^{-}$ and $H^{--} H^{+}$
provide the largest cross sections among all channels. Due to their electromagnetic charge, the cross section of pair produced doubly charged Higgs bosons is large compared to singly charged Higgs pair production. 


\begin{figure}[t]
\centering
\subfloat[]{\includegraphics[width=9.8cm]{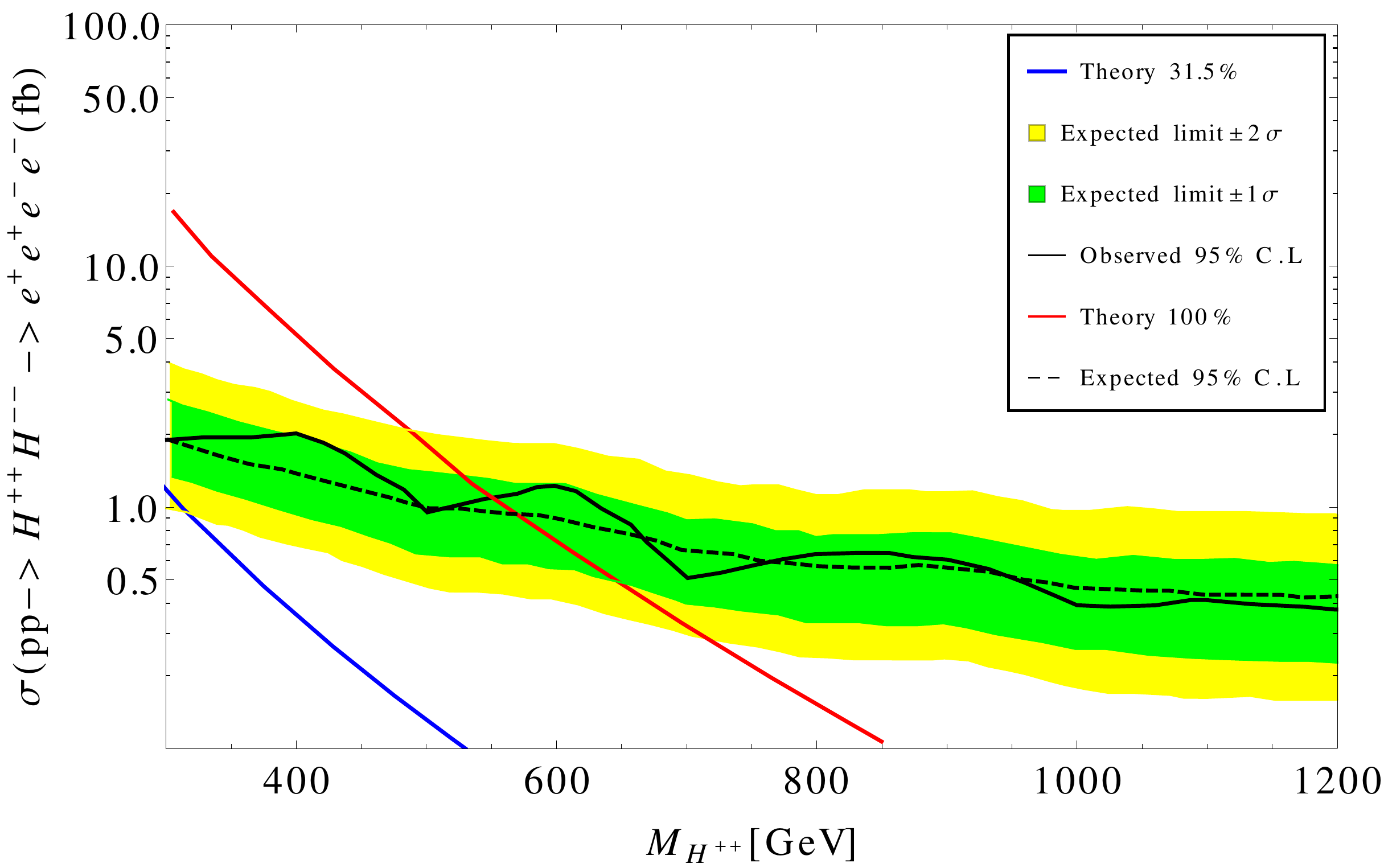}\label{}}
\caption{The limit on the cross section folded with branching ratios from 13 TeV LHC results \cite{atlashpp}. Blue line represents limit
for our scenario with $v_{\Delta}=10^{-9}$ GeV, where branching ratio of $H^{++} \to e^{+} e^{+}$ is 0.315. The $K$-factor for 13 TeV limit has been taken as 1.25 \cite{Akeroyd:2005gt}.}
\label{fig:limit13}
\end{figure}

\begin{figure}[t]
\centering
\subfloat[]{\includegraphics[width=9.5cm,height=6.5cm]{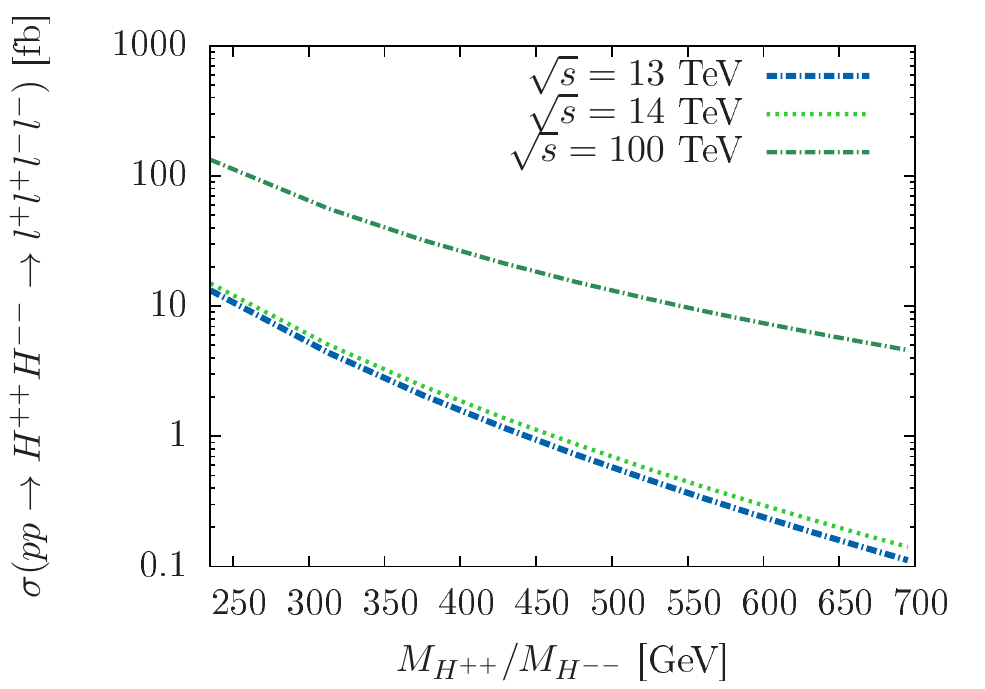}\label{}}
\caption{The cross section for $4l$ for 13, 14 and 100 TeV c.m.energy. The $K$-factor for 13 TeV limit has been taken as 1.25 \cite{Akeroyd:2005gt}.}
\label{fig:4lcross}
\end{figure}
\section{Multilepton Signature at LHC and VLHC \label{analyses}}

A large number of final states can arise from the pair production of doubly or singly charged Higgs bosons.  ATLAS 
has performed searches for doubly charged Higgs bosons in the same-sign dielectron channel for $\sqrt{s} = 13 $ TeV with $13.9\, \text{fb}^{-1}$ of data \cite{atlashpp}. 
Assuming a 100$\%$ branching ratio of $H^{\pm \pm} \to e^{\pm}e^{\pm}$, a lower bound on the doubly charged Higgs 
mass is obtained $M_{H^{++}} \ge 570$ GeV. However, depending on the parameter space, this limit can be relaxed due to the presence of other decay modes.
For illustration, we consider a scenario where $v_{\Delta} \sim 10^{-9}$ GeV resulting in BR($H^{++} \to e^{+}e^{+}$)$=0.315$. 
The bound on the mass of doubly charged Higgs bosons becomes significantly weaker, as shown  in Fig.~\ref{fig:limit13}.
The red line corresponds to the theory prediction from \cite{atlashpp}. The limit on $M_{H^{++}}$ remains mostly unchanged
as the branching ratio of the $H^{++}$ decaying into leptons are largely constant for $v_{\Delta} \lesssim 10^{-5}$ GeV (see  Fig.~\ref{fig:hpp630}). 
Instead, for larger $v_{\Delta} \gtrsim 10^{-5}$ GeV, the branching ratio into leptonic final states is even more suppressed, as the other modes, i.e. decays into gauge bosons, start to dominate and, hence, the limit becomes irrelevant. Another bound presented by CMS for the $H^{++} j j$ channel can only constrain the triplet vev  $v_{\Delta} \ge 16$ GeV \cite{Khachatryan:2014sta} which  is out of our region of interest.

Below, we consider three separate regions for the triplet vev, namely; low, intermediate and large, and discuss various multilepton signatures relevant for each. 

\subsection{Small $v_{\Delta}$ ($ \le 10^{-6}$ GeV)}

The most promising channel for small $v_{\Delta}$ is the search for pair produced $H^{++}$ decaying into dilepton. Hence, the final state consists of 4 leptons.
We show the variation of cross section of $p p \to H^{++ } H^{--} \to 4l$ with $m_{H^{\pm \pm}}$ for different c.o.m energies $\sqrt{s}=13,14$ and $100$ TeV in Fig.~\ref{fig:4lcross}. For the 13 TeV c.o.m energy the cross section is greater than 1 fb upto mass 
$M_{H^{\pm \pm}} \sim 450$ GeV. We perform a detailed signal and background analysis for the 4l case at 13 TeV c.o.m. energy at the LHC. We also show the results for 100 TeV c.o.m. energy relevant for the VLHC.

%
%
 
{\bf Analysis:} We consider two benchmark scenarios, a low and a high mass doubly charged Higgs boson, both in agreement  with the present bound from the LHC (see Fig.~\ref{fig:limit13}): 
\begin{itemize}
\item
{$M_{H^{++}} = 375 $ GeV, obtained from $v_{\Delta}=10^{-9}$ GeV and  $\mu=4 \times 10^{-9}$ GeV.}
\item
{ $M_{H^{++}} = 630 $ GeV with $v_{\Delta}=10^{-9}$ GeV and $\mu=10^{-8}$ GeV.}
\end{itemize}

\begin{figure}
\centering
\subfloat[]{\includegraphics[width=8.0cm,height=6.0cm]{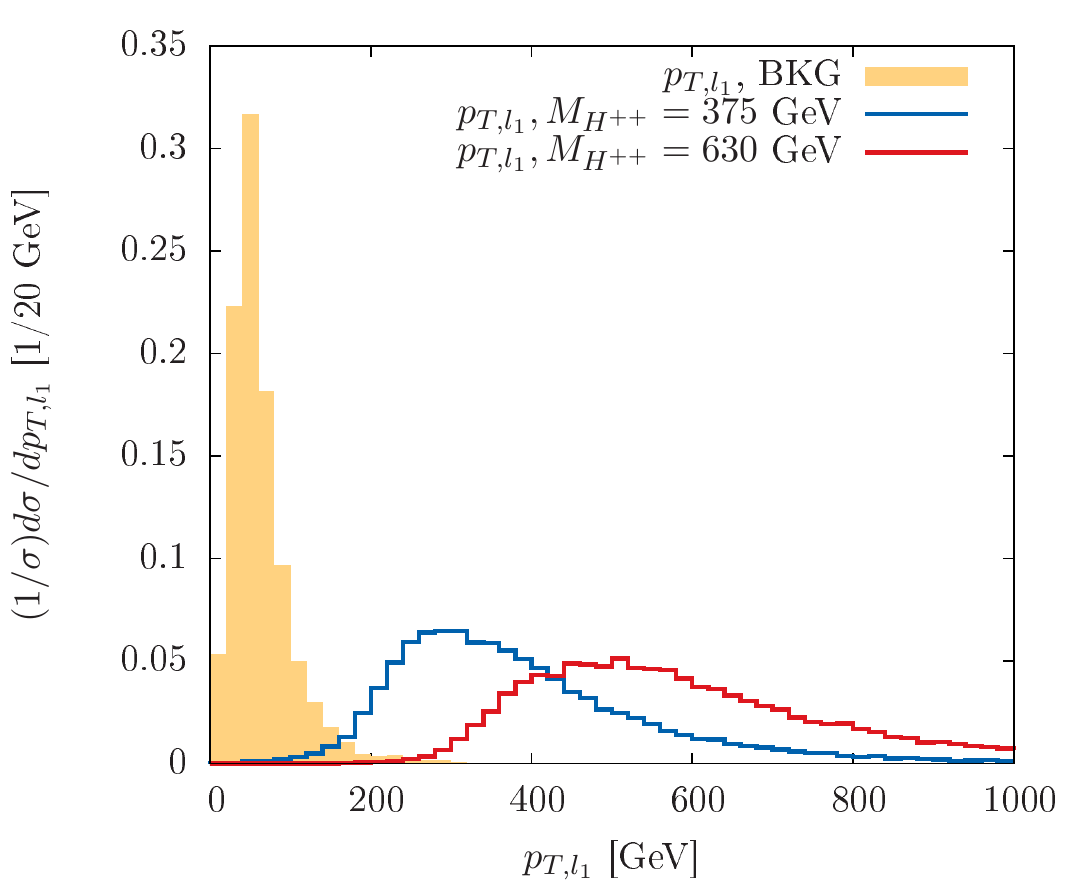}\label{fig:h4la}}
\subfloat[]{\includegraphics[width=8.0cm,height=6.0cm]{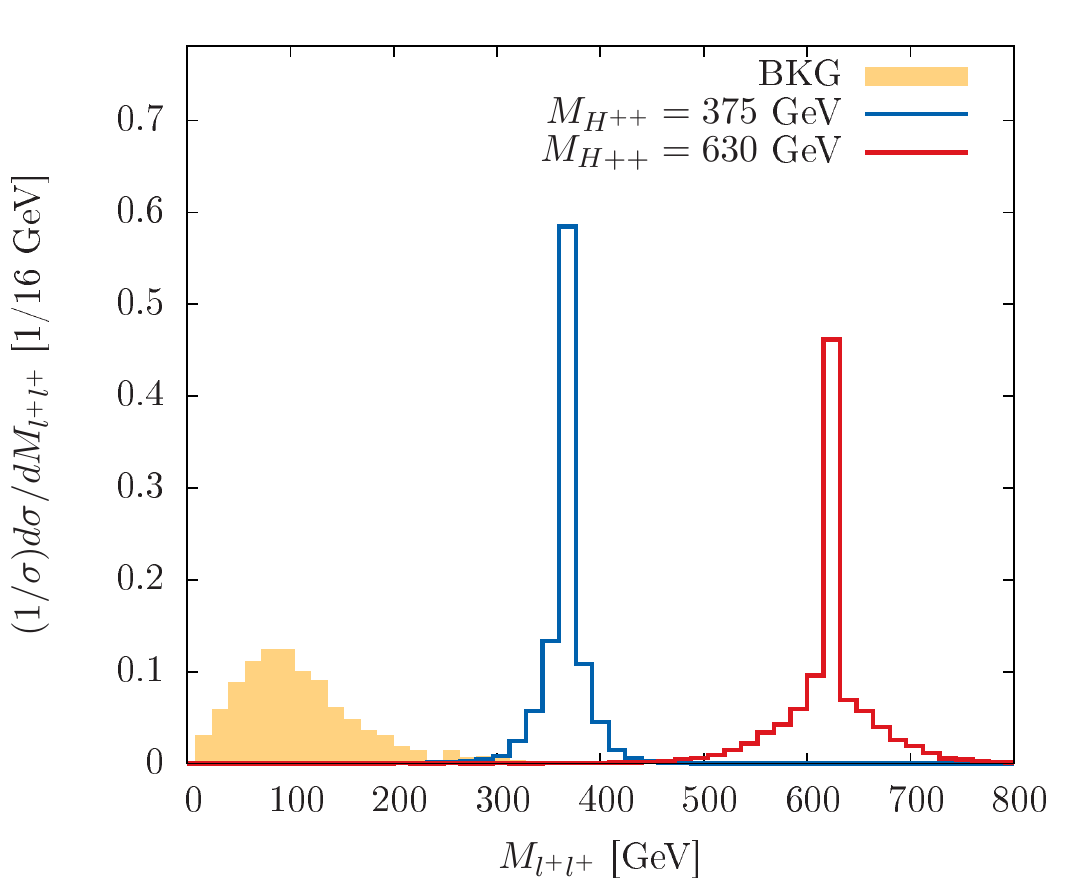}\label{fig:h4lb}}
\caption{ Left: $p_T$ distribution of the leading lepton.
          Right: Invariant mass distribution of two same-sign leptons. Both distributions correspond to the  scenario $v_{\Delta}=10^{-9}$ GeV.
          The solid curve is for the SM background. }
\label{fig:hist4l}
\end{figure}

We generate the  4l background which arises predominantly through SM diboson production. In Fig.~\ref{fig:hist4l} we show the distribution of various variables before cuts. Fig.~\ref{fig:hist4l} (left) describes the $p_T$ of the
hardest final state lepton. The invariant mass of the two positively-charged leptons $M_{l^{+} l^{+}}$ is shown in the right panel of Fig.~\ref{fig:hist4l}.
We use the following isolation and selection criteria for the final state leptons ($e^{\pm}$ and $\mu^{\pm}$):
\begin{itemize}
 \item 
  $|\eta| < 2.5$ and $p_{T,l} >  20$ GeV.
  \item
  To avoid any contamination from jet fakes, we require the hadronic activity within a cone 
  $\Delta R = 0.4$ around an isolated lepton to be $p_{T,\mathrm{had}}\le 0.15~p_{T,l}$.
\end{itemize}
We apply a series of analysis cuts in order to improve the separation of signal and background:
\begin{itemize}
\item
 a strict $p_T$ requirement for the hardest lepton: $p_{T,l_1} > 100 $ GeV.

 \item
invariant mass of the same-sign lepton pair: $|M_{l^{+} l^{+}}-M_{H^{++}}| \le 100$ GeV.
\item
veto events with invariant mass around the $Z$ peak:  $|M_{l^+l^-}-M_Z| \le 10$ GeV. 
\end{itemize}

The cross sections after analysis  cuts are given in Table.~\ref{tab:event_4l} for the two illustrative mass points.  For the lower charged Higgs mass of 375 GeV, the cross section before and after cuts are 1.659 fb and 0.827 fb, respectively. For the higher mass 630 GeV the cross sections are 0.149 fb and 0.074 fb. In addition, we also investigate the above channel for the $100$ TeV collider, where the cross-section increases by a factor $30$. 

\begin{table}[!ht]
\centering
\begin{tabular}{||c|c|c|c|c|c|c||}
\multicolumn{7}{c}{$\sqrt{s}=13$ TeV} \\
\hline 
\multicolumn{4}{|c||}{ $p p \to H^{++} H^{--} \to l^{+} l^{+} l^{-} l^{-}$} & 
\multicolumn{3}{|c||}{Background $p p \to 4l$}  \\
\hline \hline
Masses  & Cross section & Cross section & Number & Cross section & Cross section & Number \\
(GeV) & before cuts (fb) & after cuts (fb) & of events & before cuts (fb) & after cuts (fb) & of events \\
\hline \hline
375  & 1.659 & 0.827 & 248 &  51.12 & 0.0107 & 3 \\
630 & 0.149 & 0.074 & 22 & 51.12 & 0.0015 & $\sim$ 0 \\
\hline \hline
\multicolumn{7}{c}{$\sqrt{s}=100$ TeV and $\mathcal{L}=30~\mathrm{fb}^{-1}$} \\
\hline
\multicolumn{4}{|c||}{ $p p \to H^{++} H^{--} \to l^{+} l^{+} l^{-} l^{-}$} & 
\multicolumn{3}{|c||}{Background $p p \to 4l$}  \\
\hline \hline
Masses  & Cross section & Cross section & Number & Cross section & Cross section & Number \\
(GeV) & before cuts (fb) & after cuts (fb) & of events & before cuts (fb) & after cuts (fb) & of events \\
\hline \hline
375  & 32.16 & 7.66  & 229 &  335.1 & 0.057 &  1 \\
630 & 6.317 & 1.415 & 42 & 335.1 & $6.7\times 10^{-3}$ &  $\sim 0$ \\
\hline \hline
\end{tabular} 
\caption{The cross sections and the number of events after the final selection cuts for the channel $p p \to H^{++} H^{--} \to l^{+} l^{+} l^{-} l^{-}$, where $l=e, \mu$. The vacuum expectation value 
of Higgs triplet $v_{\Delta}=10^{-9}$ GeV. The number of events for both 13 TeV and 100 TeV c.o.m energy have been computed with 300 and 30 $\rm{fb}^{-1}$ luminosity. 
In the high luminosity run
of 13 TeV LHC at 3000 fb$^{-1}$, the number of observed events may increase one order higher than the
numbers in column four.
}
\label{tab:event_4l}
\end{table}


The inclusive partonic cross section for the SM background at 13 TeV is $\sim 51 $ fb, and thus much larger than the signal cross section. However, the dilepton invariant mass cut  along with the $Z$ veto are extremely helpful to reject the background. We compute the statistical significance of this channel as
\begin{equation}
n=\frac{\mathcal{S}}{\sqrt{\mathcal{S}+\mathcal{B}}},
\end{equation}
where $\mathcal{S}$ and $\mathcal{B}$ represent the number of signal and background events. We find that
\begin{itemize}
\item
The doubly charged Higgs boson of mass  $M_{H^{++}}=375$ GeV can be discovered at the LHC with 300 $\rm{fb}^{-1}$ luminosity with more than 5$\sigma$ significance, while for the higher mass of $630$ GeV, the significance is around 4.66$\sigma$ with the same amount of data.
\item
A heavier doubly charged Higgs of mass 630 GeV can be discovered at HL-LHC (13 TeV) or  VLHC (30 fb$^{-1}$) with more than 5$\sigma$ significance. 

\end{itemize}

\subsection{Intermediate $v_{\Delta}$ ($10^{-6}-10^{-2}$ GeV) \label{intvdelta}}
\begin{figure}[t]
\centering
{\includegraphics[width=9.0cm,height=7.0cm]{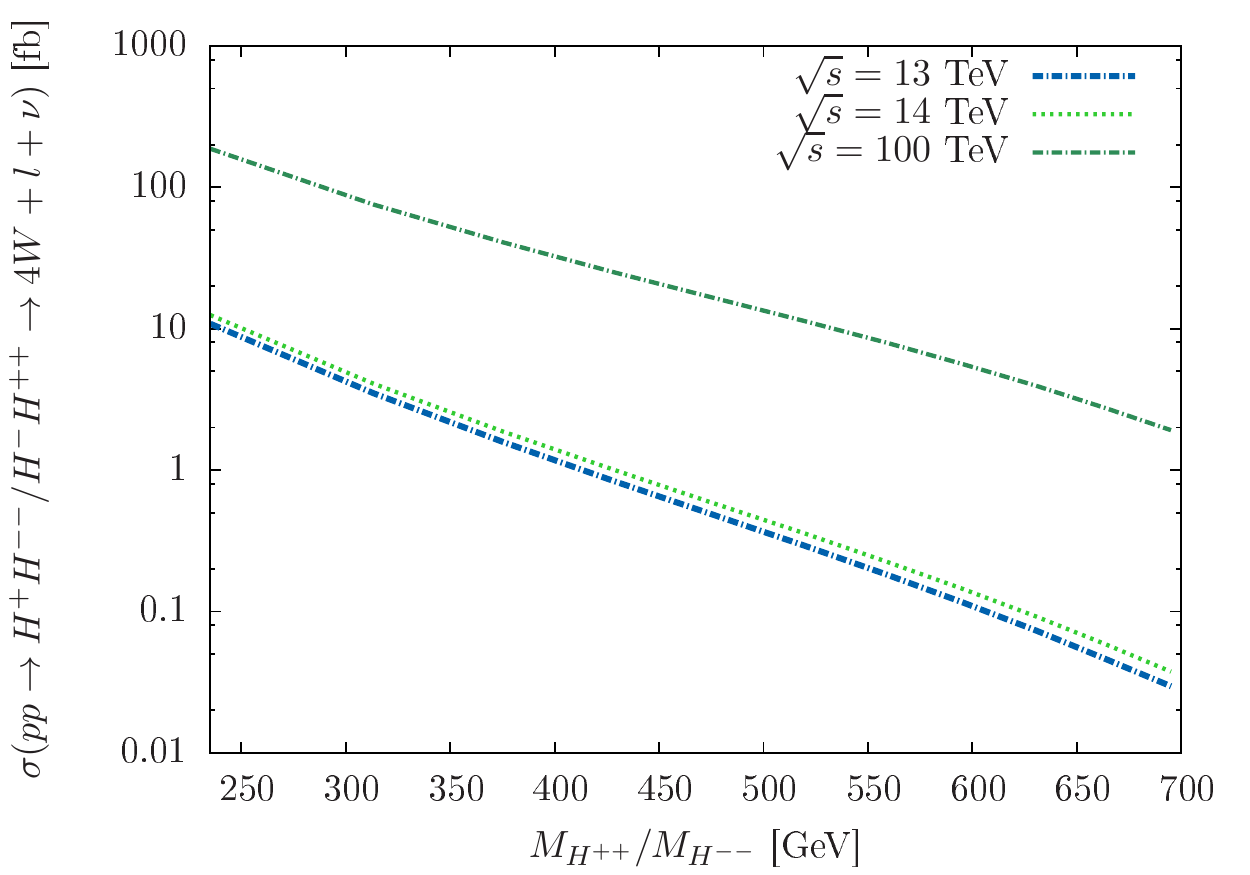}}
\caption{Cross section for the $p p \to H^{+} H^{--} \to 4 W+ l+\slashed{E}_T$ chanel via cascade
({\it i.e.}  off-shell or on-shell $W$) decay of $H^{+} \to H^{++} l \slashed{E}_T$ for $v_{\Delta}=10^{-2} $ GeV. The conjugate channel is also included. For the 13 and 14 TeV, 
we consider a K-factor $K=1.25$  \cite{Akeroyd:2005gt}.  }
\label{fig:x4w5w}
\end{figure}

In this region, $H^{+}$ preferably decays into, either on-shell or off-shell, $H^{++} W^{-}$, depending on the mass splitting between the two Higgs states. $H^{++} $ subsequently decays either into two leptons or into two $W$ bosons. The branching fraction into gauge bosons, i.e. $H^{++} \to W^+ W^+$,  becomes dominant for $v_{\Delta} > 10^{-4}$ GeV. 
This intermediate range of $v_{\Delta}$ allows for signatures with five (or even six) leptons. Below, we discuss two channels with four and five leptons in the final states
\begin{itemize}
\item
$p p \to H^{\pm \pm} H^{\mp}$, subsequently $H^{\mp} \to {W^{\pm}}^* H^{\mp \mp}/W^{+} H^{\mp \mp} \to 5W \to 5l+\slashed{E}_T$. 
\item
$p p \to H^{\pm \pm} H^{\mp}$, subsequently $H^{\mp} \to {W^{\pm}}^* H^{\mp \mp}/W^{\pm} H^{\mp \mp} \to 5W \to 4l+2j+\slashed{E}_T$. 
\end{itemize}

The large lepton multiplicity reduces the cross section, but results in a cleaner ({\it i.e.} background free) signal.
The parton level cross section for  $4W+l+\slashed{E}_T$ is shown in Fig.~\ref{fig:x4w5w}.
%
We adopt the following criteria for the leptons and jets  reconstruction:
\begin{itemize}
 \item 
  $|\eta_l| < 2.5$ and $p_{T,l} >  20$ GeV and hadronic activity around an isolated lepton within a cone of
  $\Delta R = 0.4$  has to be $p_{T,\mathrm{had}}\le 0.15~p_{T,l}$.
  \item
  $|\eta_j| < 4.7$ and $p_{T,j} >  30$ GeV.
\end{itemize}
\begin{table}[h]
\centering
\begin{tabular}{||c|c|c|c|c||}
\multicolumn{5}{c}{$\sqrt{s}=13$ TeV} \\
\hline 
Channel & $M_{H^{++}}, M_{H^{+}}$ (GeV)   &  $\sigma_{\text{before cuts}}$ & $\sigma_{\text{after cuts}}$ & No. of events \\
 & & (fb) & (fb) & at $(300, 3000)\text{fb}^{-1}$ \\
\hline 
\hline
{($5l+\slashed{E}_T$)} &   (169, 298) & 0.024  & 0.0054  & 2, 16 \\
 &  (223, 332) &  0.0124 & 0.0034    & 1, 10 \\
\hline
\hline
{($4l+2j+\slashed{E}_T$)} &   (169, 298) &  0.076  & 0.036  & 10, 107 \\
 &  (223, 332 ) &   0.0393  & 0.016     & 5, 47 \\
\hline 
\hline
\end{tabular} 
\caption{The cross sections before and after cuts for the channel $p p \to H^{\pm \pm} H^{\mp} \to H^{\pm \pm}H^{\mp\mp} W^{\pm} \to 5 W $ with $H^{+}$ decaying to on-shell $W^{}$.
Demanding $W$ to decay into leptonic or hadronic mode gives two different final states (a) All leptonic: ($5l+\slashed{E}_T$) and (b) 1 Hadronic: ($4l+2j+\slashed{E}_T$). 
The vacuum expectation value of Higgs triplet $v_{\Delta}=10^{-3}$ GeV. The number of events have been computed with 300 $\rm{fb}^{-1}$ and 3000 $\rm{fb}^{-1}$ for $\sqrt{s} = 13$ TeV at the LHC. See text for further details.  }
\label{tab:event_5lonshell}
\end{table}

In Table.~\ref{tab:event_5lonshell}, we show the cross sections for the above channels assuming on-shell decays only. Again, we consider two benchmark points with masses $M_{H^{++}}=223$ ($169$) GeV,
and $M_{H^{+}}=332$ ($298$) GeV.
In order to obtain large enough mass splittings for decays into on-shell $W$, $\lambda_4$ needs to be tuned to values $\sim \mathcal{O}(1)$.

We find that the final state with five leptons occurs only in a handful of events for an integrated luminosity of 300\, $\rm{fb}^{-1}$ at the LHC. As these processes are limited only by their rate, the HL-LHC provides a promising environment. The main  contribution for the five-leptonic states comes from triple gauge boson production with a cross section of $ \sim 0.025$ fb.
Hence, we do not analyse the background for the 5l channel. 

The dominant background for $4l + \text{jets} + \slashed{E}_T$ is $t\bar{t}Z$ with $\sigma(t\bar{t}Z) \simeq 586.4$ fb. After applying a $Z$ veto cut, the remaining cross section is reduced to a manageable rate of $0.094$ fb. 
After all cuts we find a signal cross section of $0.036$ fb for the masses $M_H^{++}= 169$ GeV and $M_{H^+}= 298$ GeV.
The signal for the above masses can be probed with a significance of $5.45\sigma$ with $3000~ \rm{fb^{-1}}$ and $1.73\sigma$ for $300~\rm{fb^{-1}}$, respectively.

\begin{figure}[t]
\centering
{\includegraphics[width=9.0cm,height=7.0cm]{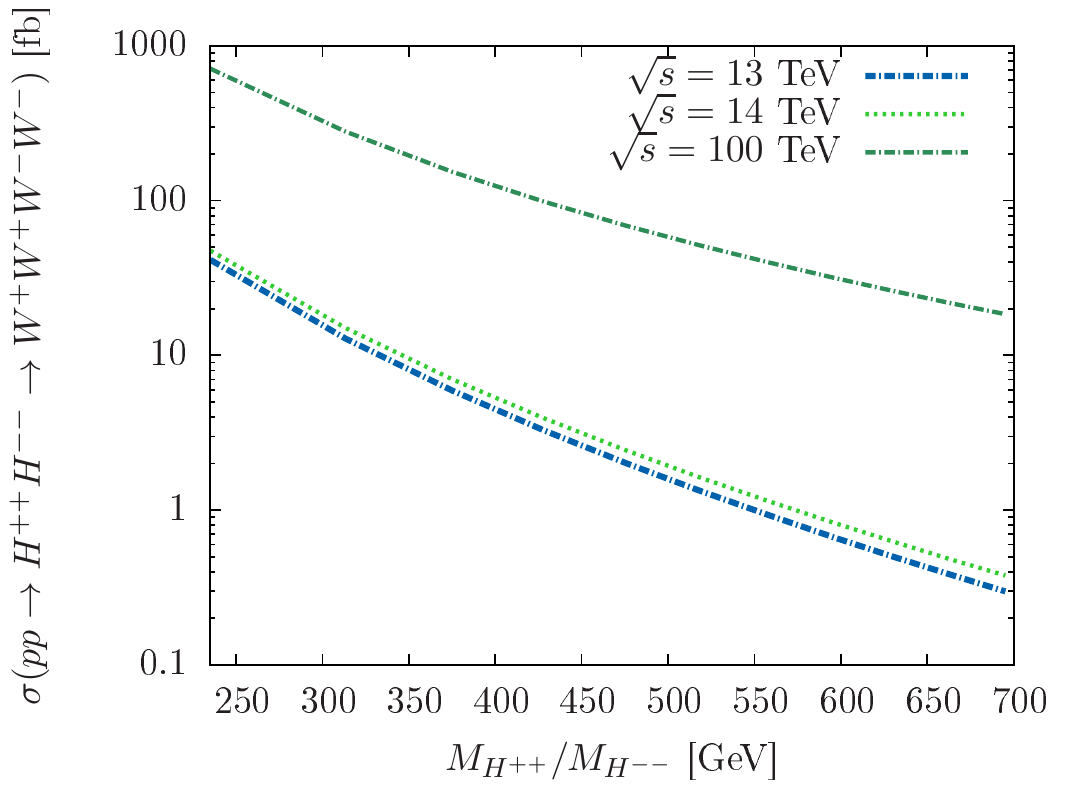}}
\caption{Cross section for the $p p \to H^{++} H^{--} \to 4 W$ against varying mass for fixed $v_{\Delta}=1 $ GeV. For the 13 (14) TeV c.m.energy, we consider $K=1.25$.}
\label{fig:x5w}
\end{figure}

\subsection{Large $v_{\Delta}$ ($> 10^{-2}$ GeV)} 
In the large $v_{\Delta}$ region ($\ge 10^{-2}$ GeV), the branching ratio of $H^{\pm \pm} \to W^{\pm} W^{\pm}$ is enhanced (see Fig.~\ref{fig:hppbr}).
We consider pair production of $H^{++}H^{--}$ where subsequent decays of the gauge bosons into leptons gives rise to the $4l + \slashed{E}_T$ signature. This has recently been analyzed in non-minimal composite Higgs scenario \cite{Englert:2016ktc}.  
Note that, this decay mode $H^{\pm \pm} \to W^{\pm} W^{\pm}$ is very poorly constrained by LHC searches \cite{Khachatryan:2014sta} and a lighter doubly charged Higgs is yet not ruled out.
We consider $v_{\Delta}=1$ GeV and $M_{H^{++}}$ as low as 235 GeV for this analysis. We show the production cross section for the process $p p \to H^{++} H^{--} \to 4 W$ in Fig.~\ref{fig:x5w}. 

The cross section of the  fully leptonic channel at 13 TeV is too small. Hence, we focus on higher c.o.m energy $\sqrt{s} =100$ TeV. This channel
can also produce a combination of leptonic and hadronic final states which will not be considered here.  We estimate the following SM processes as backgrounds:
\begin{itemize}
\item
$p p \to  4lZ$ and $Z \to \nu \nu$.
\item
$p p \to 4lW$ with subsequent decays of  $W \to l \nu$ (with a lepton escaping detection). 
\item
$p p \to 2lWW $ with subsequent decays of  $W \to l \nu$. 
\end{itemize}
In addition to the above processes, we also consider the $t \bar{t} Z$ process followed by the further decays of $t \to b W$ and $Z \to l^{+} l^{-}$, that can generate $4l+\slashed{E}_T$ associated with $b$-jets. Note that, the $b$-jet from the above mentioned background has large $p_T$. Hence, in spite of large cross section, most of this background events can be rejected applying jet and $Z$ veto.

\begin{figure}[h]
\centering
\subfloat[]{\includegraphics[width=8.5cm]{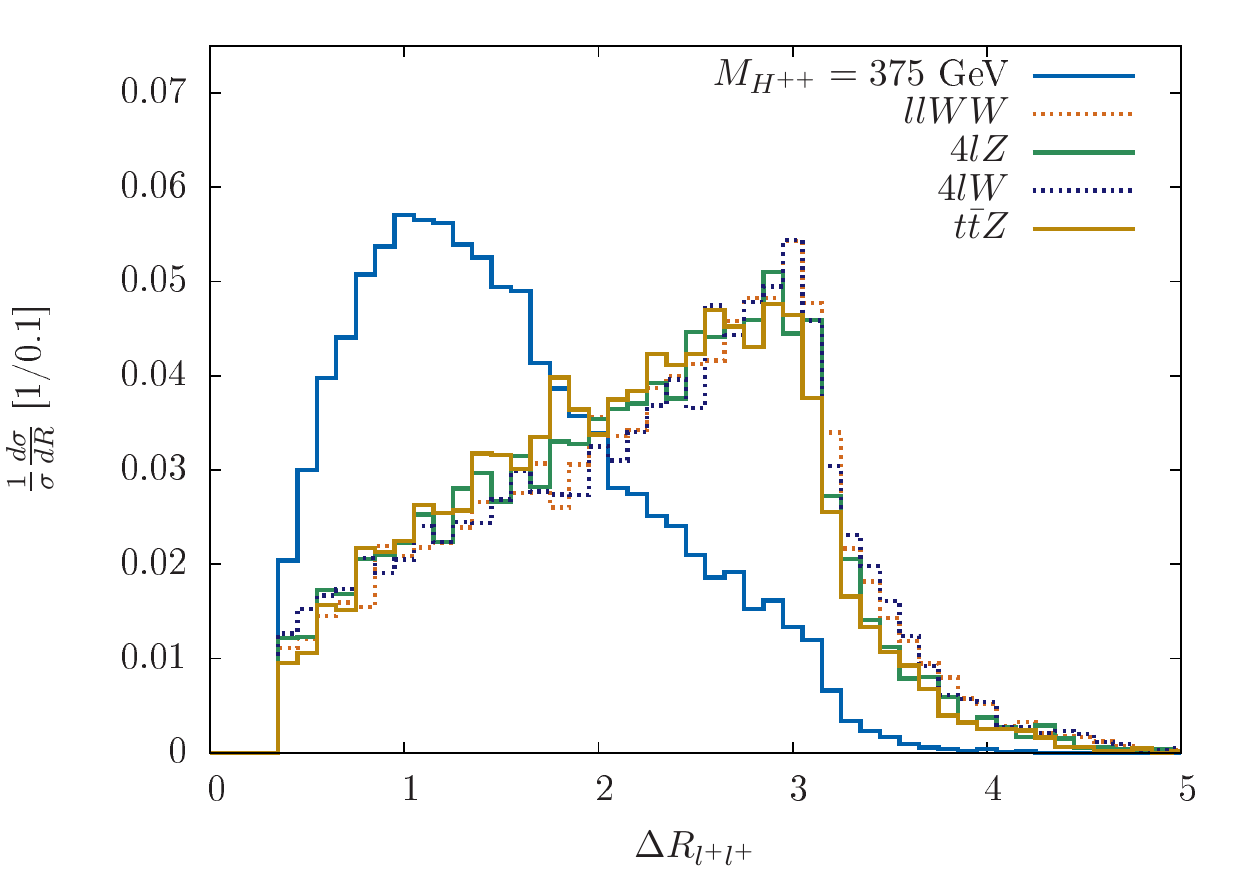}\label{fig:drpp}}~~
\caption{$\Delta R$ distribution for two same sign leptons  for the process $pp \to 4l+\slashed{E}_T$ assuming $v_{\Delta}=1$ GeV. The center-of-mass energy is $\sqrt{s}=100$ TeV.}
\label{fig:drpppt}
\end{figure}

In the signal, the $H^{\pm\pm}$ decay into a pair of $W^{\pm}$ bosons which subsequently decay into leptons, hence, final states are collimated in the lab-frame and populate the same hemisphere in the detector (see Fig.~\ref{fig:drpppt}). Such configurations are much less likely in SM processes. We exploit this by applying a cut on the $\Delta R$ separation of the same-sign dilepton system.
We demand four leptons in the final state and employ the following sets of analysis cuts:
\begin{itemize}
\item
veto events with jet leading jet $p_T > 40$ GeV,
 \item
 $\Delta R < 2.0$ between two same-sign leptons,
\item
veto events with $|M_{l^{+}l^{-}}-M_Z|<10$ GeV.
\end{itemize}

\begin{table}[!ht]
\centering
\begin{tabular}{||c|c|c|c|c||}
\multicolumn{5}{c}{$\sqrt{s}=100 ~\rm{TeV}$}\\
\hline 
Channel & $M_{H^{++}}$ (GeV)   &  $\sigma_{\text{before cuts}}$  & $\sigma_{\text{after cuts}}$ & No. of events \\
 & & (fb) & (fb)& at $300~\text{fb}^{-1}$ \\
\hline \hline
 $p p \to H^{++} H^{--} \to 2W^{+}2W^{-}$  & 235 GeV &  0.643   &  0.355    & 106 \\ 
  $\to 4l+\slashed{E}_T$ & 375 GeV & 0.155  &  0.093   & 27 \\
\hline  \hline
Background & & & & \\
\cline{1-1} 
\rm{$p p \to 4lZ, Z \to \nu \nu$} & - & 0.291  & 0.0068  & 2\\
\rm{$p p \to 2lW^{+}W^{-}, W \to l  \nu$} & - & 3.12  & 0.097  & 29\\
\rm{$p p \to 4lW, W \to l \nu$} & - & 0.31  &  0.0073  & 2\\
\rm{$p p \to t \bar{t}Z \to 4l+2b+\slashed{E}_T$} & - & 48.51  & 0.0165   & 4 \\
\hline \hline
\end{tabular} 
\caption{The cross sections at 100 TeV collider after basic trigger cut and selection cuts for the channel $p p \to H^{++} H^{--} \to W^{+} W^{+} W^{-} W^{-} \to 4l+\slashed{E}_T$, 
where $l=e, \mu$. The vacuum expectation value of Higgs triplet $v_{\Delta}=1$ GeV. The number of events have been computed with the aim of
achieving $300~\rm{fb}^{-1}$ luminosity.}
\label{tab:event_4w4l}
\end{table}

The cross sections are shown in Table.~\ref{tab:event_4w4l} for two illustrative mass points $M_{H^{++}}=235$ and $375$ GeV. We find that the above mass points can be discovered at a VLHC with the following significance:
\begin{equation}
\frac{\mathcal{S}}{\sqrt{\mathcal{S}+\mathcal{B}}}=8.87\sigma \, (3.37\sigma) \, ~{\rm{for}}~\, M_{H^{\pm \pm}}=235 \, (375) \rm{GeV}.
\end{equation}

\section{conclusion \label{conclu}}
We investigated various multilepton signatures that arise in a Type-II seesaw model with a Higgs triplet . The model contains singly charged Higgs bosons as well as doubly charged Higgs states. 
Depending on the triplet scalar vev $v_{\Delta}$, $H^{++}$ and $H^{+}$ can have a number of decay modes. We focus on three different regimes of $v_{\Delta}$ and investigate the
multilepton final states for each different regime. For small $v_{\Delta}$ ($< 10^{-6}$ GeV), $H^{++}$ prefers to decay to two same-sign leptons.
Therefore, pair production of $H^{++} H^{--}$ leads to a distinctive four leptonic signature. Assuming a 100$\%$ branching ratios of $H^{++} \to l^{+} l^{+}$, the recent LHC search have constrained 
$M_{H^{++}} > 570$ GeV. However, the limit is considerably weakened for a parameter space with lower branching ratio to leptons.
We discuss in detail the prospects to observe this mode at the current run of the LHC and also at a future hadron collider to be run at a c.o.m energy of $100$ TeV. We summarize 
our observations as follows :

\begin{itemize}
\item
The channel with four leptons, arising from $H^{\pm\pm}$ decays offers the most promising signature for small $v_\Delta$. We conclude that a doubly charged Higgs boson of mass 375 GeV can be discovered at 13 TeV LHC with  $300~\rm{fb}^{-1}$ luminosity. Higher mass range 630 GeV can further be discovered at a high-luminosity LHC or at a VLHC with 30 fb$^{-1}$ luminosity.

\item

For the intermediate $v_{\Delta}$ range, the most distinctive channel arises from cascade decays of the singly charged Higgs $H^{+} \to H^{++} W^{-}$ (both on-shell and off-shell).
Further, $H^{++}$ can decay either in dilepton or $W^{+} W^{+}$ modes. This leads to a final state consisting of $5W$.  If all W decay leptonically rates are very small, resulting in $\sim$ tens of events for $300~\rm{fb}^{-1}$ luminosity at 13 TeV. But the signal is very clean. 
We also analyze another topology with $4l+2j+\slashed{E}_T$. SM backgrounds for these channels are extremely small which makes it an interesting search strategy. A lighter doubly charged Higgs mass around 169 GeV can be conclusively discovered with more than $5 \sigma$ at high luminosity run of LHC. 

\item
Finally, the large $v_{\Delta}$ region, which is poorly constrained at the LHC seems to be the most promising channel to probe lighter doubly charged Higgs bosons at the VLHC.  
In this case, the four leptons in the final state appears from the pair production of $H^{++} H^{--}$ followed by the decay of $H^{\pm \pm} \to W^{\pm} W^{\pm}$. We find that a doubly charged Higgs of mass 235 GeV can be discovered at $\sim 8 \sigma$ significance at the VLHC. 

\end{itemize}

The properties of the SM-like Higgs boson has been quite well established by the LHC. No new physics has been observed so far, barring some initial statistical fluctuations, fuelling hope for potential signals.
Many of the new physics models, although proposed long ago, lack detailed studies covering all their parameter regions. We explored parts of the parameter space of a SM extension with a Higgs triplet which is, otherwise, difficult to probe at existing (and future) colliders. Finding a (doubly) charged Higgs boson will be an immediate proof for the existence of at least another
$SU(2)_L$ scalar multiplet. 

\section*{Acknowledgement} 

The work of MM has been supported by the Royal Society International Exchange Program and DST-INSPIRE-15-0074 grant. MM thanks  IPPP, Durham University, UK for hospitality where part of the work was being carried out. 
SN acknowledges Dr. D. S. Kothari Post Doctoral Fellowship awarded by University Grant Commission (UGC) for financial support. MM and SN thank Dr. Shankha Banerjee and Prof. C. Lester for their  invaluable inputs and Dr. Santosh Kumar Rai for providing the FeynRules model file.

\end{document}